# Bounds on the rates of growth and convergence of all physical processes

Toby Ord*

The upper limit on what is computable in our universe is unknown, but widely believed to be set by the Turing machine — with a function being physically computable if and only if it is Turing-computable. I show how this apparently mild assumption leads to generous yet binding limits on how quickly or slowly any directly measurable physical phenomenon can grow or converge — limits that are intimately connected to Radó's Busy Beaver function. I conjecture that these limits are novel physical laws governing which rates of growth and convergence are possible in our universe.

It is commonly thought that the Turing machine provides the upper limit on what is computable in our universe. This is sometimes known as the *physical Church-Turing thesis*.[1] It is an idealised claim, saying that a finite physical machine supplied with unlimited resources (energy, storage, space, time…) could eventually compute any Turing-computable function, but can never go further. The thesis makes no reference to the efficiency of the computation. So while quantum computers might be able to compute some functions faster, the functions which they can compute with unlimited resources remain those of a classical Turing machine and thus they are compatible with the thesis.

While unproven, this physical Church-Turing thesis is the default hypothesis in computer science and physics. Suggestions that it may be possible to construct machines that go beyond the Turing machine are usually considered fanciful. This tacit acceptance of a fundamental, but unproven, physical limit may in part be because its implications are somewhat esoteric — governing a realm of intangible information and only ruling out the computation of functions so obscure that they had rarely, if ever, been posed prior to the 20th Century.

---

* I'd like to thank Scott Aaronson for rekindling my interest in uncomputable rates of growth and Eric Chen and Sami Petersen for convincing me to share these ideas.

[1] This is a different claim to the (standard) *Church-Turing thesis*, which states that the set of functions that can be computed by Turing machines are precisely those that could eventually be computed by an idealised clerk with paper and pencil, following clear rules without using insight. The Church-Turing thesis is a conceptual claim, whereas the physical Church Turing thesis is an empirical claim — it would be true under some sets of physical laws and false under others. There are several different ways of making the physical Church-Turing thesis precise, but no universally agreed formulation (Kreisel 1965, Gandy 1980, Deutsch 1985, Copeland 1997 & 2020, Ord 2002).

The most well-known function that a Turing machine cannot compute is the *halting function*. It takes a representation of a Turing machine and its input, returning 1 if the machine would eventually halt given that input and 0 otherwise.[2] The physical Church-Turing thesis implies that there is no physical system that computes this function.

This physical prohibition on computation of certain functions gives rise to prohibitions on other kinds of physical phenomena, such as patterns that can never be physically manifested. For example, a black and white linear pattern where all positions are black by default, but where position $2^n \times 3^m$ is white if the $n$th Turing machine halts on the $m$th input. By observing as much of this pattern as needed, we could compute the halting function. So the physical Church-Turing thesis says that it cannot be physically constructed — not as a pattern in space, nor as one in time (such as in the intermittent flashing of a light).

Ruling out particular physical patterns is perhaps of more interest than ruling out particular computations, since patterns are somewhat closer to the standard objects of study in the natural sciences. But this is still a very specific and esoteric pattern.

However, the physical Church-Turing thesis also has unexpectedly bold implications that fall squarely in the traditional purview of physics.

Rates of growth and decline are everywhere in physics. Some natural processes grow quickly — the distance a ball falls grows quadratically with time; populations grow exponentially. Some natural processes decline quickly — the light from a globe fades as the inverse square of the distance; the amount of radioactive material remaining in a sample declines exponentially.

If the physical Church-Turing thesis is true, then there are hard limits on how quickly (or slowly) *any* physical process can grow or decline — four hitherto unknown iron laws that govern the rates of growth and convergence that can exist in our universe.

---

[2] Equivalent uncomputable halting functions exist for all models of computation and programming languages with the same power as Turing machines. The proof is simple. If such a function were computable, one could make a new function that checked if it was going to halt (when given itself as input), then proceeded to do the opposite of what was predicted. But this would be paradoxical, so we have a proof by contradiction that a Turing-complete system cannot compute the halting function.

**The Busy Beaver function**

In 1962, Tibor Radó introduced the *Busy Beaver* function, $BB(n)$.[3] It takes a natural number $n$ as input and returns the longest finite running time of any (binary) Turing machine with $n$ states, starting with a blank input tape.[4] We can immediately see that this function cannot be Turing-computable, as if we had access to it, we could compute the halting function. To determine if a Turing machine halts on some input, we would consider a modified machine that first writes that particular input onto the blank tape. For an input of length $l$, the modified machine would have $l+n$ states. So we could simulate it for $BB(l+n)$ steps, knowing that if it hasn't halted by then it will never halt, and thus that the original Turing machine wouldn't halt on that input.

While the Busy Beaver function is canonically defined on Turing machines, one can define a version for any Turing-complete model of computation or programming language. For example, $BB_{python}(n)$ is the largest number of timesteps a python program of length $n$ characters with empty input can run before halting. All such Busy Beaver functions grow astonishingly quickly, and all share analogous interesting properties.

One of the most interesting of these is that the Busy Beaver function marks a line in the sand where every single function which grows as fast or faster is also uncomputable by Turing machines. Let's say that the function $f(x)$ *grows faster than* $g(x)$ (in symbols: $f(x) \succ g(x)$) if and only if there is some number $K$ such that for all $x > K$, $f(x) > g(x)$. If a function, $f(x)$, were to grow faster than the Busy Beaver function, a computer could use it to compute the halting function. We could just use the previous method but replace our upper bound for how long an $n$-state machine could run before halting with $\max(f(K), f(n))$, which is always greater than $BB(n)$.[5]

We can even go a little further: Radó (1962) proved that every Turing-computable function $f(n)$ *grows slower* than the Busy Beaver function. That is, $BB(n) \succ f(n)$. This rules out the possibility that a Turing-computable function might neither surpass, nor be surpassed by, the Busy Beaver function — that they might keep swapping the lead infinitely often. Such rates of growth wouldn't always allow one to solve the halting problem but are also out of reach for Turing machines.

---

[3] See Aaronson (2020) for a good recent survey of what is known about the behaviour of the Busy Beaver function.

[4] In his original paper, Radó introduces two functions $\Sigma(n)$ and $S(n)$. $S(n)$ is the one I discuss in this paper, whereas $\Sigma(n)$ is the longest number of consecutive 1s an $n$-state binary Turing machine can output when starting with a blank tape. The term 'Busy Beaver' was initially associated with $\Sigma(n)$ but is mostly now associated with $S(n)$. Either function would suffice for this paper's purposes.

[5] Not knowing when the function overtakes $BB(n)$ — thus not knowing the value of $K$ — would be an impediment to practically using this, but wouldn't undermine the theoretical point about the limits of Turing machines or the limits of any physical machines. The physical Church-Turing thesis claims that there is no way of configuring the universe to compute functions that aren't Turing-computable. Whether or not we'd recognise that configuration is beside the point.

And it is not just that Turing-computable functions grow slower than the Busy Beaver function — they can't even get close. They grow slower than $BB(n)/1000$ and $\log(BB(n))$ and even $\log(\log(\ldots\log((BB(n))\ldots))$ with $n$ nested logarithms. They can't even get into the same ballpark as the Busy Beaver function, let alone beyond.

The Busy Beaver function grows very quickly. Its first four values are known precisely: 1, 6, 21, 107. One can see from this that the function is accelerating, but there is little indication of just what is to come. $BB(5)$ is 47,176,870. And $BB(6)$ has been proven[6] to exceed:

$$10^{10^{10^{10^{10^{10^{10^{10^{10^{10^{10^{10^{10^{10^{10}}}}}}}}}}}}}}$$

The Busy Beaver function grows faster than any polynomial or exponential; faster than iterated exponentials or the famed Ackermann function. To my knowledge, it grows faster than any function that had been proposed in the history of mathematics prior to 1962.

**Limits on the growth of physical phenomena**

If true, the physical Church-Turing thesis would turn mathematical results like these into physical laws. It implies that no physical computer — however it was constructed — could compute a function that grows faster than the Busy Beaver function. And because any physical process with a directly measurable quantity that grows faster than the Busy Beaver function could be used to physically compute such a function, the physical Church-Turing thesis rules them out too:

> No physically realisable process (starting with a finite configuration plus unlimited resources and time) can contain a directly measurable quantity that grows faster than the Busy Beaver function.

The Busy Beaver function would thus act as a kind of cosmic speed limit on observable rates of growth. Just as the speed of light is a hard limit on physical speeds, so the Busy Beaver function would be a hard limit on growth of any directly measurable physical quantity.[7] Because the Busy Beaver function grows unbelievably

[6] Ligocki (2022).

[7] Note that there do exist specifiable physical quantities that grow faster than the Busy Beaver function — but they would never be directly measurable. As a simple example: the amount of

quickly, it is a very generous limit. But then so is the speed of light. They provide for an extremely generous range of possible speeds or rates of growth, but then absolutely rule out those that go any further.

One might think that these rates of growth are so preposterous that relativity or quantum mechanics already rules them out. For example, if it was a distance that was growing faster than the Busy Beaver function, it would eventually have to grow faster than the speed of light.[8] Or if it was a mass that was growing quickly, it would eventually be limited by the rate at which new matter or energy could be found to feed the process — and even bringing in new resources at the speed of light would only allow cubic growth over the long run.

But there are other ways of setting things up that wouldn't obviously violate any known physical laws. For instance, we could imagine a system that alternates ever more slowly between two states, with the lengths of time between alternations growing faster than the Busy Beaver function. This need not involve unlimited speed, mass, or energy, nor arbitrarily accurate measurement. But it could be used to solve the halting problem, so would be ruled out by the physical Church-Turing thesis.

And such an alternating system would allow one to compute the halting function even if the system couldn't be made perfectly precise or reliable. For example suppose it occasionally failed to switch states. If one was reading off the duration of the $n$th period between alternations and using that as an upper bound for how long a Turing machine with $n$ states could run before halting, then occasional failure to switch would increase the length of that period between alternations. And it would also mean that each subsequent period between alternations happened when the count of alternations was one fewer — effectively translating the rest of the graph of the function one step to the left and thus *increasing* its rate of growth. So the computational payload of such a high rate of growth would be robust to this kind of error. Therefore even an unreliable system of this kind would be ruled out by the physical Church-Turing thesis.

---

time that a particular physically specified computer system can run before halting given any input of size $n$. If this machine acted as a universal Turing machine, simulating the machine that it takes as input, then the longest time it could run grows as the Busy Beaver function itself. But this time is not at all a directly measurable physical quantity of that system. Indeed, determining this quantity by examining the system would require additional computation beyond that of the Turing machine.

[8] Even this isn't expressly ruled out by general relativity. For instance, the current default model of cosmology (ΛCDM) involves space itself expanding at a rate that approaches exponential. The distance between here and a distant galaxy eventually grows faster than the speed of light, though not in a way that allows super-luminal information transfer.

So far, we've focused on rates of growth of some measurable physical quantity over time. But the limit applies more broadly. If any directly measurable quantity in a system increased as a function of any other directly measurable quantity faster than the Busy Beaver function one could also compute the halting function. So the physical Church-Turing thesis also sets hard limits on how quickly size can increase as a function of mass, or how quickly durations can increase as a function of distance — or any other combination of measurable quantities.

**A plethora of physical limits**

So far, we've shown how the Busy Beaver function can act as an upper bound on rates of growth. But that's not the only kind of bound the physical Church-Turing thesis sets on growth or decline. There is an entire menagerie of related bounds.

To the Busy Beaver, who works harder and harder so quickly that we cannot overestimate its future industriousness, we could add the Sleepy Sloth with its supernatural capability to wait ever-more preposterously long periods of time before getting the next thing done. Then there is the Dawdling Daydreamer who in their extreme reluctance to finally arrive at their destination, approaches it more and more painstakingly slowly, and finally Asymptoting Achilles who excels at approaching his goal at an ever-more blinding rate without ever actually arriving.

We can define the *Sleepy Sloth* function as $SS(n) = BB^{-1}(n)$. That is, it is the inverse of the Busy Beaver function — its reflection around the diagonal. $SS(1) = 1$, and so does $SS(2)$ through $SS(5)$. Only at $SS(6)$ does it rise to 2. The horizontal pauses between each rise grow longer and longer. Even $SS(47{,}176{,}869)$ is still 4. The sleepy sloth function grows to infinity but does so unbelievably slowly. Far slower than square root, or log, or any monotonic Turing-computable function.

If you had any physical system with a directly measurable quantity that grew slower than the Sleepy Sloth function, you could compute the halting function — just invert it to find a function that grows faster than the Busy Beaver function. Indeed we have already encountered this function in disguise: the slowly increasing count of alternations in the hypothetical physical system described earlier was growing over time as the Sleepy Sloth function.

The *Dawdling Daydreamer* function is defined by $DD(n) = 1/SS(n)$. This function approaches zero, but like its namesake does so unbelievably slowly, closing in by smaller and smaller fractions of the remaining distance. Any Turing-computable process that declined more slowly than this would stall out at some finite distance away from zero. If you had a physical quantity that truly approached zero (or any other number) more slowly than this, you would be able to compute the halting function.

And finally, the *Asymptoting Achilles* function is defined as $AA(n) = 1/BB(n)$. It also approaches zero but closes in by greater and greater fractions of the remaining distance each time step. No Turing-computable function can approach zero (or any other number) faster than the Asymptoting Achilles function without actually getting there.

These can be further generalised. No directly measurable quantity could grow to minus infinity faster than $-BB(n)$ or slower than $-SS(n)$. No directly measurable quantity could converge to some finite level *from below* faster than $-AA(n)$ or slower than $-DD(n)$. Even if the convergence involves subsequences converging from above and below, both subsequences have their rate of convergence bounded by these functions, and the absolute discrepancies between points in the sequence and its limit are bounded by $AA(n)$ and $DD(n)$. Even the rate of convergence towards an arbitrary computable function $f(n)$ would be bounded — the absolute discrepancies between the physical values and $f(n)$ can't decline faster than $AA(n)$ or slower than $DD(n)$.

While these functions are formally defined only on a discrete countable set of inputs, they can also be extended to continuous versions over the rationals or reals. There is no canonical way of doing so, but any monotonic extension (even just connecting successive points with diagonal line segments) will give a continuous function with the same key properties — acting as bounds for the relevant rates of growth or convergence.

The bounds relating to convergence are physically relevant only if the quantity on the $x$-axis is measurable to an arbitrarily fine degree. For many physical quantities, issues of quantisation or measurement in quantum mechanics may already rule out gradual convergence at any rate whatsoever. But these proposed bounds would also rule out convergence in quantities where arbitrary precision has not yet been ruled out. For example, they would bound the rates at which *averages* can converge, such as the average spin of all particles emitted so far from a particular machine of arbitrary construction. Measuring this average to arbitrary precision doesn't appear to run into any established quantum limits.

The physical Church-Turing thesis thus imposes four novel limits on the rate at which any directly measurable physical quantities can grow or converge (see *Figure 1*). None can grow towards infinity (without reaching it) faster than the Busy Beaver function or slower than the Sleepy Sloth function. None can approach a finite limit (without reaching it) faster than the Asymptoting Achilles function or slower than the Dawdling Daydreamer function. Indeed, they can't even get within any constant factor of any of these limits of growth or convergence.

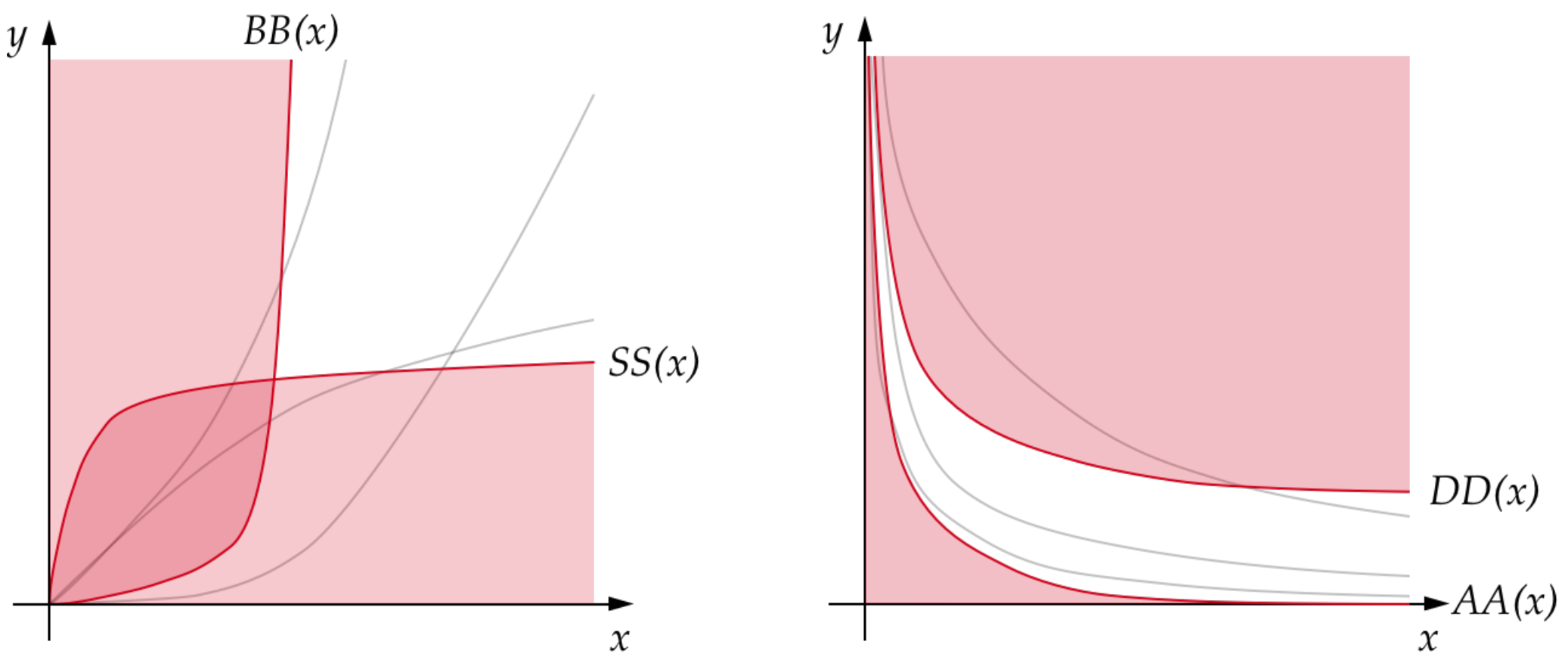

***Figure 1.*** *Physically possible rates of growth and convergence over an infinite interval.* If the physical Church-Turing thesis is true, all physically possible rates of growth and convergence must eventually leave the shaded regions and remain in the white region thereafter. The faint grey curves are examples of rates of growth or convergence that are compatible with the thesis. On the left are the rates of growth to infinity in infinite time, on the right the rates of convergence to a finite value in infinite time. Note that the curves shown are schematic and not to scale.

For the sake of completeness, we could also note four further limits that govern rates of growth and convergence over a finite interval. First consider functions that grow monotonically to infinity with a vertical asymptote at $x = k$. None can grow faster than $DD^{-1}(k–x)$ or slower than $AA^{-1}(k–x)$. More precisely, for every directly measurable physical quantity that grows to infinity at $x = k$, there must be some $\epsilon$ such that when $x$ is within $\epsilon$ of reaching $k$, the quantity is lower than $DD^{-1}(k–x)$ and higher than $AA^{-1}(k–x)$.

Or consider functions that converge towards a finite value, reaching it at precisely the point $x = k$. None can converge in on it faster than $1/DD^{-1}(k–x)$ or slower than $1/AA^{-1}(k–x)$ (see *Figure* 2).

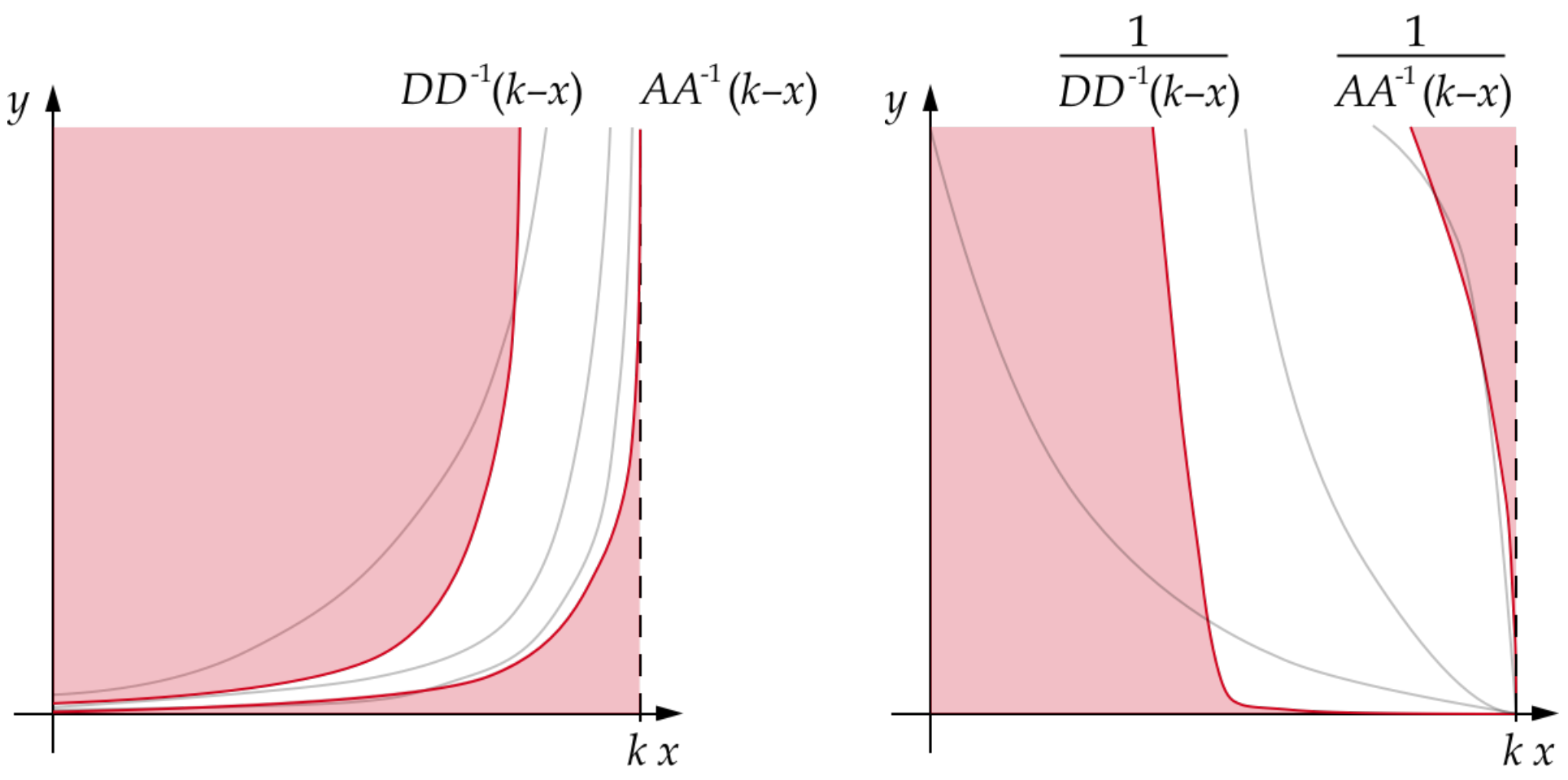

***Figure 2.*** *Physically possible rates of growth and convergence within a finite interval.* As per Figure 1, but now these diagrams describe the limits on growth to infinity in finite time (Left) or convergence to a finite value in finite time (Right).

It is worth reflecting on just how generous all these bounds are. Many physical bounds restrict possibilities that one might have actually wanted to achieve or expected to observe in natural phenomena. That is not the case here. Very few people would have even considered processes that grow or converge quickly (or slowly) enough to be forbidden by these bounds. At best, this is a case of not knowing you wanted something until the moment you are told it is impossible.

Furthermore, all of these involve the asymptotic behaviour of things that grow arbitrarily large or converge arbitrarily closely to some value. And it may be that our actual universe doesn't allow *any* directly measurable quantity to grow arbitrarily large or to converge arbitrarily closely without reaching its limit. Even true linear growth or exponential decay would be ruled out. This could be the case if the space of possibilities within the affectable part of the universe is fundamentally finite. If so, then additional limits on how quickly or slowly infinite processes might grow or converge, would add very little as a practical matter.

That said, even if the space of possibilities is fundamentally finite, the study of asymptotics or limiting behaviour of physical processes is still a core part of theoretical physics. The particular finite bounds would be contingent, and we could always ask how things would have proceeded had they been larger and larger. This is the context in which the physical Church-Turing thesis is usually presented: there may be constraints (such as a finite amount of reachable matter) that prevent anyone building even a full Turing machine, but there is still considerable interest in considering what would be possible given enough resources.

It is in this theoretical context that these new physical limits are of interest. Not that they put much constraint on what we are likely to find or build in our universe, but through the very idea that rates of growth or convergence are the kinds of things that could even *have* upper and lower bounds set by the laws of physics. In this way, it is perhaps less interesting *what* these limits are, and more interesting *that* there are any such limits at all, *why* theories of computation necessitate such limits, and *how* they forge a connection between computability and physical growth.

**Generalised Busy Beavers**

We have so far been considering the Busy Beaver function (and friends) for Turing machines and other computationally equivalent machines or programming languages. There is also a tradition going back to Turing himself, of considering even more powerful computational systems. Turing (1939) put this in terms of a Turing machine that had access to an *oracle* — some special device that could directly return the results of a specific function that need not be Turing computable. For example, one can consider what a Turing machine could compute given access to an oracle for the halting function and even develop algorithms that show how it could use this to compute a host of other functions.

Computability theory has explored these ideas deeply, finding transfinite hierarchies of functions that are increasingly difficult to compute (Odifreddi 1989). For example, it is always impossible for an oracle machine to compute the halting function for oracle machines that use that very same oracle function. So if we were to consider regular Turing machines to be 'level zero', we could define level $n$+1 machines to be oracle machines with an oracle for the halting function for level $n$ machines. In this way, we can stratify the computational difficulty of many functions by which level machine is required to compute them.[9]

Just as a machine can't compute the halting function for machines of its own kind, it can't compute their Busy Beaver function. If we consider the set of all oracle machines with a given oracle function, $o(\cdot)$, we can define a Busy Beaver function for them, $BB_o(n)$. We just set $BB_o(n)$ to the largest (finite) number of steps an oracle machine with oracle $o(\cdot)$ can run before halting, when given a blank tape as input. Small tweaks to the standard proofs show that $BB_o(n)$ will grow faster than any function computable with oracle $o(\cdot)$. So, for example, $BB_h(n)$ where $h(\cdot)$ is the halting function, grows much faster than $BB(n)$. The hierarchies of ever-more powerful oracle machines can thus be used to create a hierarchy of ever-faster-growing Busy Beaver functions (Aaronson 2020). And each of these Busy Beaver functions comes with its own set of reflections and reciprocals that bound all kinds of growth and convergence for any function computable using a given oracle.

This implies that even if the physical Church-Turing thesis is *false* — and computation beyond that of the Turing machine is possible in our universe — so long as this computation equivalent to (or weaker than) any kind of oracle machine, all directly measurable physical quantities would *still* have their growth and convergence bounded by a set of eight functions analogous to those described above. However, if we don't know in advance what functions are computable in our universe, we won't be able to specify which set of eight Busy Beaver–like functions act as these bounds.

**Bounds on oscillation and more**

These ideas can be further extended to place limits on many other ways that physical quantities can change. For example, consider smooth functions for which these limits are inapplicable because they neither grow without bound, nor converge to a particular value. Such functions must have a kind of oscillatory behaviour with infinitely many turning points (that need not be periodic). The physical Church-Turing thesis also places limits on how these turning points can be arranged.

[9] This infinite hierarchy and its transfinite extensions are just one part of the structure of levels of computability. The full structure involves parts that are dense (with levels of computability between any two levels) and incompletely ordered.

For example, while the frequency of turning points can rapidly increase (such as in the function $\sin(e^x)$), it can't grow towards infinity faster than the Busy Beaver function or more slowly than the Sleepy Sloth. Moreover, the gaps between successive turning points can't grow in length more quickly than the Busy Beaver function or more slowly than the Sleepy Sloth function. Indeed there are versions of all eight functions that bound what can happen to the pattern of turning points in any directly physically measurable quantity.[10]

And the same is true for other countably infinite sets of distinguished points on a function — such as $x$-intercepts, vertical asymptotes, points of inflection, or cusps. So long as their locations can be physically determined, the rates at which they appear as $x$ increases are all limited by the eight functions in the Busy Beaver family.

**Conclusions**

The primary result of this paper is that the physical Church-Turing thesis implies a specific set of generous — but unyielding — bounds on how quickly or slowly any directly measurable physical quantity can grow or converge, based on reflections and reciprocals of the (standard) Busy Beaver function.

We could understand this implication in several different ways.

Since the physical Church-Turing thesis is indeed a very plausible and widely held assumption, these new bounds on growth and convergence could be seen as interesting new physical laws.

Alternatively, one could see this result as pushing back on the physical Church-Turing thesis. We've perhaps been too quick to accept it, and now that we realise it implies hitherto unstated laws of physics that put hard limits on all forms of growth and convergence, it might be time to reconsider. That said, the existence of generalised Busy Beaver functions suggests that some hard bounds on growth and convergence of physical phenomena are likely to be required regardless, so rejecting the physical Church-Turing thesis on these grounds may be a hollow victory.

Finally, we could interpret the implication as a more neutral disjunction: either the widely accepted physical Church-Turing thesis is in fact false, or there are these new physical laws. An interesting result either way (though it would be nice to know which!).

Personally, I conjecture that these novel physical laws about how quickly or slowly measurable quantities can grow or converge really do govern our universe. They aren't yet proven, so for now they are merely conjectural bounds. (I further conjecture that if growth isn't bounded by the Busy Beaver function itself, it will be bounded by some higher-order Busy Beaver function, leading to similar laws.)

[10] Even if the frequency of turning points neither grows to infinity nor converges to some fixed value, it may *still* be subject to related bounds. In this case, the frequency would *itself* be oscillatory and thus subject to higher order versions of these limits.

Even the standard Busy Beaver function produces a set of bounds so generous that they are unlikely to provide any practical hindrance to our goals or constraint on what we expect to find in the natural world. Their chief interest instead lies in the very idea that physical rates of growth can be (and perhaps are) subject to a well specified upper limit, and that this provides a linkage between computability and physical rates of change.